\documentclass[11pt,letterpaper]{article}

\usepackage{hyperref}
\usepackage[numbers]{natbib}
\usepackage{xspace}
\usepackage{xcolor}
\usepackage{graphicx}
\usepackage{amsmath,amssymb}
\usepackage[margin=1in]{geometry}
\usepackage{tikz}
\usetikzlibrary{chains,arrows}
\usepackage{todonotes}

\newtheorem{proposition}{Proposition}
\newtheorem{theorem}{Theorem}[section]
\newtheorem{lemma}[theorem]{Lemma}

\def\squareforqed{\hbox{\rule{2.5mm}{2.5mm}}}

\def\QED{\ifmmode\squareforqed 
  \else{\nobreak\hfil   
    \penalty50                 
    \hskip1em                  
    \null                      
    \nobreak                   
    \hfil                      
    \squareforqed              
    \parfillskip=0pt           
    \finalhyphendemerits=0     
    \endgraf}                  
  \fi}

\def\blksquare{\rule{2mm}{2mm}}
\def\qedsymbol{\blksquare}
\newcommand{\bg}[1]{\medskip\noindent{\bf #1}}
\newcommand{\ed}{{\hfill\qedsymbol}\medskip}

\newenvironment{proof}[1][]{{\it{Proof #1. }}}{\ed}

\newcommand{\FullversionOmit}[1]{} 
\newtheorem{claim}[theorem]{Claim}

\newcommand{\fullversion}[1]{}

\usepackage{color}

\usepackage{pgfplots}
\pgfplotsset{compat=1.3}

\pgfplotsset{
	tick label style = {font=\scriptsize},
	every axis label = {font=\scriptsize},
	legend style = {font=\scriptsize},
	label style = {font=\scriptsize},
	/pgfplots/xlabel near ticks/.style={
		/pgfplots/every axis x label/.style={
			at={(ticklabel cs:0.5)},anchor=near ticklabel
		}
	},
	/pgfplots/ylabel near ticks/.style={
		
		/pgfplots/every axis y label/.style={
			at={(ticklabel cs:0.5)},rotate=90,anchor=near ticklabel}
	}
	
}

\begin{document}

\title{Effect of selfish choices in deferred acceptance with short lists}
\author{Hedyeh Beyhaghi
\thanks{{\tt hedyeh@cs.cornell.edu}, Dept of Computer
Science, Cornell University. Supported by ONR grant N00014-08-1-0031.}
\and
Daniela Saban
\thanks{ {\tt dsaban@stanford.edu}, Graduate School of Business, Stanford University.}
\and \'Eva Tardos
\thanks{ {\tt eva@cs.cornell.edu}, Dept of Computer
Science, Cornell University. Work supported in part by NSF grant CCF-1563714, ONR grant N00014-08-1-0031, and a Google faculty research award.}
}
\date{}
\maketitle

\begin{abstract}
We study the outcome of deferred acceptance when prospective medical residents can only apply to a limited set of hospitals. This limitation requires residents to make a strategic choice about the quality of hospitals they apply to. Through a mix of theoretical and experimental results, we study the effect of this strategic choice on the preferences submitted by participants, as well as on the overall welfare.  We find that residents' choices in our model mimic the behavior observed in real systems where individuals apply to a mix of positions consisting  mostly of places where they are reasonably likely to get accepted, as well as a few ``reach'' applications to hospitals of very high quality, and a few ``safe'' applications to hospitals of lower than their expected level. Surprisingly, the number of such ``safe'' applications is not monotone in the number of allowed applications. We also find that selfish behavior can hurt social welfare, but the deterioration of overall welfare is very minimal.
\end{abstract}

\section{Introduction}
We consider a two-sided matching market where individuals apply to positions and monetary transfers are not permitted. The main goal in such markets is typically to find a stable matching, i.e. a matching in which no pair of agents would rather be matched with each other than with their current partners.
Given the preference lists of the applicants and the positions, a stable matching can easily be found using the much celebrated Gale-Shapley  mechanism \cite{Gale1962}, which constitutes the basis for many real-world centralized matching mechanisms. Examples include the National Residency Matching Program (NRMP) used to match medical students with residency programs, public schools assignments in the US, and the assignment of seats in public universities in other countries (see e.g. \cite{Roth1992}).

An important feature of the markets described above, which is usually not modeled by the classical work studying the Gale-Shapley mechanism, is that the number of applications sent by an individual is in general much lower than the number of open positions. For example, in school and university assignment systems there is often a constraint on the number of applications that can be submitted. Even if no such formal constraint is imposed, applications still might require a fair amount of work (e.g university applications typically require a separate essay for each school) or might impose a time constraint (e.g. in the NRMP, interviews must be conducted before the application process) which then naturally limits the number of applications submitted per individual.

The goal of our paper is to study the incentive properties of the the Gale-Shapley mechanism in systems with limited number of applications. It is well known that in the Gale-Shapley stable matching mechanism with men proposing to women, it is dominant strategy for the men to report their preferences truthfully \cite{Roth1982}, while women can benefit from strategic behavior \cite{Gale1985}. However, the truthful reporting for men is relying on the fact that they can report their complete preference list. 
When the number of applications is limited, the mechanism is no longer truthful and deciding where to apply to becomes non-trivial.
%

To illustrate, consider the college application process where options on either side of the market are not ex-ante identical: preferences of applicants and schools are highly correlated ---not surprisingly, applicants generally prefer good schools, and all schools  want to fill their positions with the best applicants. 
If the student has a low SAT score, he will very likely not get into a top-tier school. 
However, while college application tools and high school councilors try to help students understand the level of school they should expect to be admitted at, many low performing students still choose to apply to one (or a couple) of their dream schools. Is this behavior rational? Similarly, top-students with almost perfect SATs tend to add a couple of safe choices to their application lists. In any case, applicants usually do not (and should not) simply truncate their true preference lists, and have to employ more complex strategies instead. The question then becomes: how should applicants decide where to apply in matching markets with short lists, when agents are not ex-ante identical and preferences are correlated?

The objective of this paper is three-fold: understand how strategic applicants decide where to apply, how these decisions are affected by the market primitives and, what is the ``quality" of the outcome in these markets.

\paragraph{\textbf{Our Results.}}
We consider a simple model that allows us to study the questions proposed above. We assume that both sides of the market (referred to as doctors and hospitals hereafter) are divided into tiers. For simplicity, we assume two tiers on each side, high and low, and the number of doctors and hospitals to be equal in each tier, although some of our results can be extended to more tiers and unequal sides.\footnote{In particular, our equilibrium characterization is also valid if the number of agents on each side is not equal.} To model correlations in preferences, we assume that all agents agree that the  high-tier doctors are better than the low-tier ones, and top tier hospitals are better than lower tier ones, but have heterogeneous within-tier preferences which we assume to be drawn uniformly at random. With our assumptions, the Gale-Shapley algorithm with full preference lists would match high doctors to high hospitals and low doctors to low hospitals, so we think of this assignment as the reasonable expectation of the doctors.

We study the mechanism in which each doctor submits a list of $K$ applications, for some parameter $K$ of the problem. Each hospital ranks all doctors, and we run the doctor-proposing Gale-Shapley algorithm.
The utility that an agent (hospital or resident) obtains from a match is driven primarily by the tier of his partner. With preferences within each tier uniformly random, applicants simply need to decide how many hospitals in each tier to list.\footnote{It is easy to see that within each tier he will list his top choices. We defer the discussion to the Section~\ref{sec:model}, once the model is formally introduced.}

\paragraph{Characterization of Nash Equilibrium.} To understand how applications are decided in this model, we use the large market approximation introduced in \cite{Arnosti2015}. {Our first main result is Theorem \ref{thm:equilibrium_two_tiers}, showing} that there exists a unique symmetric Nash equilibrium of this game, which has a nice structure: either there is a pure-strategy equilibrium, or an equilibrium in mixed strategies in which doctors randomize between ``close" configurations (there exists $k_h, k_l$ in $\{0, \ldots, K-1\}$ such that doctors in tier $i \in \{h,l\}$ randomize between listing $k_i$ or $k_i+1$ high hospitals and the rest low hospitals.)
This equilibrium captures the main behavior observed in practice: while most applications are devoted to the most reasonable alternatives, top doctors still list some safe hospitals, while bottom doctors list a couple of ``reach'' hospitals.

\paragraph{Worst case welfare properties.} {We measure welfare of an assignment of doctors to hospitals by considering the value both for doctors as well as for hospitals. Unassigned doctors and unfilled hospital positions derive no value. We assume that the value for a doctor of being assigned to a low hospital is 1, and the value for being assigned to a high hospital is $v$. While doctors have strict preferences among individual hospitals within a tier, we assume that these differences in value are small relative to the overall quality of the hospital; i.e. what tier they belong to. Similarly, we use 1 as the value for a hospital for being assigned a low doctor and $v$ for the value for a high doctor.}

Our second main result, Theorem \ref{thm:K>1}, shows that the social welfare produced by the Nash equilibrium is always within a factor of 2 of the optimal welfare. {This low price of anarchy is surprising, as  selfish behavior of the doctors (of applying to too many ''reach'' hospitals), is depriving some of the lower hospitals from any doctors, and hence can do significant damage to welfare. In fact,} if we allow for the number of high hospitals to be (much) smaller than the number of top quality applicants, the Nash equilibrium  produces significantly lower welfare than that of the social optimum; in such case, the welfare loss  is no longer bounded by a constant and rather becomes a function of the ratio of applicants to hospitals.
To understand the effect of applications to ``safe'' and ``reach'' schools, we also compare the welfare of the Nash equilibrium against that produced by a simple market design, in which doctors can only apply to hospitals in their own tier. {It is not hard to see that the welfare of this simple market design is within a $e/(e-1)$ factor of the optimum. We show that} the Nash outcome can be either worse or better than this simple design in terms of social welfare, it is at most a $e/2(e-1)\approx 0.79$ factor worse than this simple restricted solution.

\paragraph{Sensitivity to the market design.} Through a combination of theory results and simulations, we study how the primitives of the market impact the strategies used by doctors, and the total welfare that is produced.
First, we show that the number of high applications $k_h$ submitted by high doctors is monotone in $K$, but 
the number of applications sent to ``safe'' hospitals can {change in surprising ways, e.g.,  can} decrease when increasing number of overall applications. {We give an example, where with $K=5$ applications high doctors apply each to 1 high hospital and 4 low hospitals, while with $K=6$ applications they apply to 3 hospitals of each type.}
We find empirically that high doctors typically apply to at most 1-2 low hospitals (``safe'' options), while low doctors can be applying to many more high hospitals.
We also study the effect of increasing $K$ on welfare empirically for various values for high match, and ratios of low and high hospitals and applicants.
	
%
%
\paragraph{\textbf{Related Literature.}}
To the best of our knowledge, the first paper that studied incentives in Gale Shapley mechanism with short preference lists is Immorlica and Mahdian \cite{ImmorlicaMahdian2005}. They show that, when doctors have short lists, with high probability the stable matching is unique and truth-telling is the best response for hospitals. Arnosti \cite{Arnosti2015} studies the quality of the outcomes in terms of the welfare under different (hospital) preference structures. However, both papers assume that the short preference lists of applicants describe their true full preferences, thus ignoring the strategic aspects on how doctors choose what hospitals to list.

The 
literature of college admissions also considers the issue of students being able to apply to a select set of schools.
\cite{Hafalir2014} and {\cite{Chade2014}}, study the problem with two colleges (similar to our model of two tiers of hospitals) and heterogeneous students who must decide {which of the two schools they apply to}. 
Avery and Levin \cite{Avery2010} study a game theoretic model of early decision systems used in US universities, where students can apply early to a single school, either just signaling their preference, or committing to the school if admitted. 
{These} papers consider the decentralized problem where not only doctors are strategic on where to apply, but also hospitals are strategic on setting their admissions standards or early admission policies. {In contrast,} 
we consider a centralized system but each tier has multiple hospitals, and
we allow doctors to submit a list containing more than one (or two) hospitals. While this adds realism to the model and the insights obtained, it also adds significant complications to the analysis.

In the papers mentioned so far, as well as in our work, the applicants know their own preferences and decide to apply to hospitals or schools aiming to maximize the quality of the school they get accepted at. Kadam \cite{Kadam2015}, Drummond et al \cite{Drummond2016}, and Kleinberg et al \cite{Kleinberg2016} study models where costly exploration is necessary to discover the applicant's ranking. Due to the discovery cost, these  models also lead to exploring (and then ranking) only a limited set of options. Drummond et al \cite{Drummond2016} and Kadam \cite{Kadam2015}, study a two-stage model, in which doctors choose a short application set and interview at this set of hospitals. In both cases, they consider markets where both doctors and hospitals are (almost perfectly) vertically differentiated. On the other hand, we consider a model where many ex-ante identical students (those in the same tier) might be competing for the same school. Also, we provide bounds on the overall welfare implication of the system design. 

\section{Model}\label{sec:model}
We consider a two-sided market consisting of doctors and open positions in hospitals (hospitals for short). For simplicity, we assume there are $N$ doctors and $N$ positions, and that each hospital has only one open position, so we focus on one-to-one matching. Each agent has a full ordered list of preferences for agents on the other side of the market. Doctors, however, can only submit to the system  a preference list of length $K$, for some parameter $K$ of the system. Hospitals, on the other hand, submit a list ranking all doctors.

Once doctors and hospitals submit their preferences, a doctor-proposing deferred acceptance algorithm is run to determine the final assignment of doctors to hospitals. The algorithm starts with all doctors unmatched. In each step, {all unmatched doctors, who have not yet exhausted their options, apply to their most preferred hospital among those to which he/she has not yet applied. Now each hospital tentatively accepts their most preferred doctor from the doctors now applying and the one who has been tentatively assigned to the selected hospital, and rejects all other applicants. }
The procedure is repeated until all unmatched doctors have been rejected from every hospital in their lists. 
Naturally, the algorithm uses the lists submitted by the doctors (not their true preferences) and the hospitals' true preferences over doctors.

To model the preferences of the agents, we assume that each side is divided into two tiers, high and low. These tiers allow us to divide the agents into types, which we use to capture a vertical component in preferences by assuming that all doctors prefer any high tier  hospital to every low tier one, and similarly all hospitals prefer any high quality doctor to any low quality ones.
Under this assumption, agents' true preferences rank first all agents in the top tier, then all agents in the low tier. Within each tier, we assume that the preferences are drawn independently and uniformly at random.
Having only two tiers on each side allows us to simplify the analysis and the exposition, while clearly allowing us to distinguish between doctors applying to ``reach'', in level, and ``safe'' hospitals, a key feature of our model.\footnote{{When a low quality doctor chooses to apply to a high hospital, we view this as a \emph{``reach''} application, and when a high quality doctor applies to a low hospital, we view this as a \emph{''safe''} application.}
	} {However, the arguments used to characterize the equilibrium can be to extended to multiple tiers.}
For most of the paper, we will assume that there are $n$ high quality doctors and $rn$ low quality doctors, and $n$ high hospitals and $rn$ low hospitals\footnote{When $rn$ is not a integer number, we just approximate it to the closest integer}, with $N=n+rn$. Under this assumption, the Gale-Shapley matching algorithm with full preference lists will match high doctors to high hospitals, and low doctors to low hospitals.\footnote{This assumption is not used in the equilibrium characterization in Section~\ref{sec:eq}.}

We consider the game when doctors are allowed to list only up to $K$ hospitals, for parameters $K<<n$. As argued in the introduction, in general doctors will not want to simply list their top $K$ choices of hospitals; high doctors might prefer to list a safe option to avoid being unmatched, while low doctors who also prefer the high hospitals, should know that they are unlikely to obtain such a match. A natural strategy for each doctor would be to apply to the top $K$ hospitals within its own tier: top $K$ high hospitals for high quality doctors, and top $K$ high hospitals for low quality doctors. We call this the \emph{simple application system}. Since doctors are making offers, and hence the algorithm is truthful for them, doctors should order the same-tier hospitals using their true preference, and the only strategic decision they need to make is selecting the hospitals to apply to.\footnote{{It is not hard to extend the result of Immorlica and Mahdian \cite{ImmorlicaMahdian2005} to our tiered model to show }
that with large $n$ and constant $K$ algorithm is also strategy proof for hospitals with high probability, so thus we assume that hospitals report their preferences truthfully.}

The question we want to study is how many reach or safe hospitals do doctors apply to, and what is the effect of this selfish choice on the quality of Nash equilibria in terms of social welfare. To study these questions, we must assign values to the agent's matches. We want to focus on the low and high distinction, and hence will use the following very simple approximation for values.
The value that a doctor (from either tier) derives from getting matched to a low hospital is $1$, and the value of assigned to a high tier hospitals is $v>1$. For simplicity, we have assumed that the value is given by the tier and is independent of the actual rank of the partner in his preference list{, that is, we assumed that the difference in value within each tier is negligible compared to the difference in the value of the tiers}.
Similarly, the value for a hospital (from either tier) to be assigned a low quality doctor is 1, while the value for a hospital  to be assigned a high quality doctor is $v$, again independent of the actual rank on the hospital's preference list. {We define the \emph{social welfare} to be the sum of values obtained by all agents in the market.} Under these assumptions, it is easy to see that the maximum achievable social welfare is $2vn+2rn$, $2v$ for matching a high doctor to a high hospital and $2$ for matching a low doctor to a low hospital.
{Despite its simplicity, this model allows us to capture correlation in preferences through the tier structure and, unlike previous work, we also allow for horizontal differentiation through the within-tier heterogeneity.}

\paragraph{\textbf{Equilibrium analysis.}}

We study a ``large market approximation'' of this model, as introduced by Arnosti \cite{Arnosti2015}. In particular, we study the outcomes in the limiting case where $n$ grows (i.e. the numbers of doctors and hospitals grow), while the ratio of doctors to hospitals, the proportion of doctors and hospitals in each tier, the lengths of doctors' lists, and the valuations for each hospital/doctor are held fixed. For example, when $K=1$ and we consider the simple application system of all doctors applying to the top hospital in their own tier, we obtain that an $(1-1/e)\approx  0.63$ fraction of the doctors will be matched to hospitals.\footnote{{To see why, note that a high hospital will hire its best applicant, so the only hospitals unmatched are those with no application; this occurs with probability $(1-1/n)^n$ using the fact that the choice of all applicants is a random hospital in their tier. The approximation then follows as $(1-1/n)^n \to 1/e$ as $n \to +\infty$.}}

When $K>1$ ---doctors apply to more than one hospital--- as $n$ grows we can approximate the probability that a doctor is accepted at a hospital as fixed probability, independent of previous applications. Given the tiered structure of the hospital's preferences, we can think about the final allocation resulting from two separate steps, where first the allocation for high-type doctors is finalized and, once this is fixed, the low-type doctors are allocated. Then, focusing on the high tier of doctors,  sending some $k$ applications to high tier hospitals, the above approximations 
the probability $p$ that one single application is accepted {satisfies the following}
fixed point equation.
\begin{proposition} \cite{Arnosti2015} Using the above large  market approximation with each high doctor applying to $k$ high hospitals, the probability of a single application getting accepted is defined by equation:
\begin{equation}\label{eq:prob-p}
    (1-p)^k=e^{-(1-(1-p)^k)/p}
 \end{equation}
\end{proposition}
\proof Since each application is accepted independently with probability $p$, the probability that a doctor with $k$ applications eventually gets matched is $(1-(1-p)^k)$, and so the expected expected number of hospitals a single doctor makes an offer to throughout the Gale-Shapley process to is
    $$
    1+(1-p)+(1-p)^2+...+(1-p)^{k-1}=\frac{1}{p}(1-(1-p)^k)
    $$

Each hospital remains matched if it receives any applications. Using the approximation that all these offers are independent, the probability that a hospital didn't get an offer is then
    $$
    (1-1/n)^{\frac{n}{p}(1-(1-p)^k)}\approx e^{-(1-(1-p)^k)/p}
    $$
and so the expected number of matched hospitals is then $ n(1- e^{-(1-(1-p)^k)/p})$.
The equation claimed by the lemma then follows, as the number of matched hospitals is the same as the number of matched doctors.
\endproof

Analogous formulas also apply when the number of doctors and hospitals is not the same on the two sides of the market. {Furthermore, these formulas can also be extended to the case where different doctors apply to a different number of hospitals; this becomes useful when, for example, doctors use a mixed-strategy.
	In addition, similar formulas allow us to derive the probability that an application from a low-type doctor is accepted. In that case, we need to take into account that, every time a low doctor applies to a hospital, there exist some probability that the hospital is already taken by a high-doctor, and thus  his application will be automatically rejected due to the tiered structure in preferences. We denote the probability that a high and low hospital is already taken by $(1-\alpha_H)$ and $(1-\alpha_L)$ respectively. We make use of these large market approximations  when deriving the results in this paper.

\section{Existence and Structure of Equilibrium}\label{sec:eq}

We call an equilibrium \emph{symmetric} if all high type, as well as all low type doctors use the same strategy, while different types are allowed to have different strategies. As discussed {above,} 
the strategic decision of a doctor can be thought of as an ordered pair $(k, K-k)$, where the first (resp. second) component indicates how many high-type (resp. low type) hospitals he/she is ranking.

\begin{theorem}\label{thm:equilibrium_two_tiers} In a market with two tiers, a symmetric equilibrium always exists and is unique, and for each type of agents $i\in \{L,H\}$, the equilibrium strategy is either pure, or is a randomization between consecutive strategies $(k_i, K-k_i)$ and $(k_i+1, K-k_i-1)$ for some $k_i\in \{0, \ldots, K\}$.
\end{theorem}

{We provide a proof sketch here for high doctors only, which one can think of separately due to the tiered structure of the hospital's preferences.} The proof for the low doctors is the same, except that it must account that some hospitals are already taken by a high doctor (in that case, the low doctor gets automatically rejected). We divide the proof into two lemmas, first about the structure, and then stating the existence of equilibria. We give a short sketch of their proofs. We include full details of the proof in Appendix \ref{sec:app_eq}.

\begin{lemma}
\label{lem:structure_two_tiers}  If a symmetric equilibrium exists for high doctors, it is either an equilibrium in pure strategies, or it is a mixed strategy equilibrium where agents randomize between consecutive strategies $(k, K-k)$ and $(k+1, K-k-1)$.
\end{lemma}
\proof[Sketch]
Consider a symmetric equilibrium. Under the large market approximation discussed in the Section \ref{sec:model}, each application of a doctor to a high hospital is accepted with some probability $p$, and an application to a low hospital is accepted with a different probability $p'$.  Given the values $p$ and $p'$ we can express the utility of a high doctor as a function $f(k)$ of the number of applications to high hospitals. For example $f(K)=v(1-(1-p)^K)$ as $(1-p)^K$ is the probability under this model that none of his $K$ applications are accepted.

Given $p$ and $p'$,  the best response of a single agent is to maximize $f(k)$ over integers. (By the large market assumption, $p$ and $p'$ remain unchanged regardless of the agent's action.) The main observation is that the function $f(k)$, when viewed as a  function of a real variable $k$, is strictly concave (see the Appendix for details). This implies that the maximum over integers is either a single integer or two neighboring integer values.
\endproof

To simplify notation, for a symmetric strategy of high doctors let $X$ be the expected number of high application of a doctor, and $k=\lfloor X \rfloor$. Let $p(X)$ (resp. $p'(X)$) be the probability that a single application of a doctor is accepted by an individual high (resp. low) hospital he applies to, when the strategy is given by $X$. These probabilities are implicitly defined by equations analogous to Equation (\ref{eq:prob-p}). Despite this implicit definition, one can prove that $p(X)$ is continuous and monotone decreasing, and $p'(X)$ is continuous and monotone increasing. 
Intuitively, the more slots used to list high hospitals, the smaller the probability that a single application succeeds in getting accepted, and this relation is continuous.

\begin{lemma}
\label{lem:existence_two_tiers} A symmetric equilibrium for high doctors must exist.
\end{lemma}
\proof[Sketch]
The idea of the proof is to express the equilibrium condition using $p(X)$ and $p'(X)$ defined above. If $X$ is an equilibrium, and $k=\lfloor X \rfloor \neq X$, then doctors should be neutral between applying to either $k$ or $k+1$ high-type hospitals. Consider the possible $k+1$st application sent to high hospitals. At this point, the choice of sending one more applications to high hospitals, versus sending all remaining $K-k$ applications to low hospitals must have equal value.
\begin{equation}\label{eq:equilibrium}
p(X)v+(1-p(X))(1-(1-p'(X))^{K-k-1})=(1-(1-p'(X))^{K-k})
\end{equation}
It is not hard to see, similar to the proof of Lemma \ref{lem:structure_two_tiers}, that if the above equation holds, then $X$ defines an equilibrium.

The idea of the proof is to consider the function $g(X)$, which is the difference of the two sides of the above equation. We have that $g(0)>0$ and $g(K)<0$, $g$ is continuous at all non-integer points, and is strictly decreasing. So there is either an $X$ with $g(X)=0$, or an integer point $k$ where $g(X)$ {changes sign}. 
We show that in the latter case, $k$ defines a pure strategy equilibrium.
\endproof



%

\section{Equilibrium Efficiency}\label{sec:comp}
To study the efficiency of equilibrium, we can consider two different benchmarks:
\begin{itemize}
\setlength{\itemsep}{0pt}\setlength{\parsep}{0pt}\setlength{\parskip}{0pt}
\item Full list optimum, or {\sc optimum}: the matching resulting from running the Gale-Shapley algorithm when doctors have a full list. In this case, everybody is matched to an agent in his own tier. Noting that the  contribution to social welfare of a match on top is $2v$ and of a match in the low tier is $2$, the total welfare of this benchmark is
$2(v+r)n$, independent of $K$.\footnote{This benchmark is overestimating the actual welfare, as it assumes that doctors can rank all hospitals. While it will be fairer to compare the Nash welfare to that achieved by the optimum under short lists, we use full-list optimum as it is easier to compute.}

\item {\sc simple}: the matching resulting of every doctor sending all applications to hospitals in their own tier. We {focus on the even simpler} 
    benchmark of simple with $K=1$,
    which  makes the computations smoother, referred hereafter as {\sc simple}.
\end{itemize}

In the {\sc simple} benchmark (with $K=1$), the probability that a hospital is open in high tier is $(1-1/n)^n \approx 1/e$, similarly also $1/e$ in low tier. Therefore with large number of doctors, there are $n\frac{e-1}{e}$ matchings in the high tier and $rn\frac{e-1}{e}$ in the low tier with the total welfare $2(v+r)n\frac{e-1}{e}$. For simplicity of computation, we assume $v \ge e/(e-1) \approx 1.58$; this assumption is somewhat reasonable as the tiers represent vertically differentiation and hence there should be a non-negligible difference between high and low tiers.

We first  consider the case where $K=1$, as it allows for closed-form computations as opposed to relaying on the fixed-point equations. We then generalize our bound for $K>1$.

\paragraph{\textbf{Equilibrium when $K=1$.}} We now provide a closed-form characterization of the equilibrium when $K=1$ and $v \ge e/(e-1)$. First, high doctors apply to a high hospital with probability 1, as even when all high doctors apply to high hospitals, the probability that an application sent by a doctor is accepted is  $(e-1)/e$ which produces an expected value of $v(e-1)/e>1$.
Now assume low tier doctors send an application to a high tier hospital with probability $x$, then the expected number of applications to top is $xrn$ (with a very small error term for large values of $n$). {Now the probability that a high hospital accepts a low doctor comes out to be} $(1-e^{-xr})n/(exrn)=(1-e^{-xr})/(erx)$. Also, all $rn$ low hospitals are available and so the probability of an application to such a hospital being accepted is $(1-e^{-(1-x)})/(1-x)$. This is a Nash equilibrium if either (1) $x=1$ and  $v(1-e^{-r})/(er)\ge1$, or (2) $x=0$ and $v/e \le (1-1/e)$, or (3) $0<x<1$ and
\begin{equation} \label{eq:v}
\frac{v(1-e^{xr})}{exr}=\frac{1-e^{-(1-x)}}{1-x}.
\end{equation}
The social welfare is $2vn(1-1/e)+n(v+1)(1-e^{-xr})/e+2rn(1-e^{-(1-x)})$.

Note that even the special case with $K=1$ already exhibits some of the issues we discussed in the introduction. While high doctors apply to hospitals in their own tier, low type doctors already have incentives to apply to a ``reach'' (high) hospital {with some probability} if $v>e-1\approx 1.72$.

\paragraph{Bounds on the efficiency of the equilibrium.}
%
It is natural to conjecture that welfare of the Nash equilibrium should be at least as good as that of  {\sc simple}, as allowing low quality doctors to target high tier hospitals has significant potential to improve welfare {, even though it can also hurt welfare by leaving some low hospitals unoccupied}. However, we found that this is not necessarily the case even for large $v$.\footnote{Examples are provided in Table~\ref{tb:simpleVSnash} in the appendix.} 
Two forces can contribute to this phenomena. Doctors applying to top-tier hospitals create a negative externality for each-other by increasing the congestion, and thus decreasing the probability that other applicants get accepted. Second, the low-tier doctors trade-off only the value they expect to obtain by getting an appointment in hospitals of different tiers ($v$ versus $1$). However, the trade-off for social welfare is $v+1$ versus $2$, as the welfare of the hospitals also needs to be accounted for. Although the gap can exist, in the examples we found the decrease in welfare to be minimal. Next we show that this is also true in the worst case.\\
\begin{proposition}\label{lem:nash1}
The social welfare in Nash equilibrium in the case when $v\geq e/(e-1)$ and $K=1$ is at least a $\frac{e}{2(e-1)} > 0.79$ fraction of the welfare of {\sc simple}.
\end{proposition}
\proof[Sketch]
By our previous discussion, when $v > e/(e-1)$ every high doctor applies to high hospital in equilibrium; thus, the welfare produced by the high doctors in the equilibrium and {\sc simple} agrees. Next, 
using equation~\ref{eq:v} we can find some $v$ such that strategy $x$ is an  equilibrium for low doctors. Our goal is to find the equilibrium with highest loss of welfare. We show this happens when $x$ converges to $1$ and the efficiency in this case is more than $\frac{e}{2(e-1)}$ fraction of {\sc simple}. The intuition behind is that being strategic makes agents to apply more to top than what would be optimal in terms of welfare: 
by matching to a high hospital they obtain value $v$ versus $1$ for matching low,  while the social welfare the contribution of those matches is $v+1$ and $2$ respectively.
\endproof

We next leverage the previous result to obtain a bound on the welfare of the Nash equilibrium for the general case with $K \ge 1$ applications.

\begin{theorem}\label{thm:K>1}
{When $v\geq e/(e-1)$,} the loss of efficiency of  equilibrium when doctors have list of  size $ K \geq 1 $ compared to {\sc optimum} is at most a factor of $2$.
\end{theorem}

\paragraph{Remark.} Note that compared to the optimal welfare with full preference lists, the equilibrium welfare can be as
low as only a $(1-1/e)\approx 0.63$ fraction: this is the case with a single tier when $K=1$,  as we expect random applications to miss a $1/e$ fraction of hospitals.

\proof[sketch of Theorem \ref{thm:K>1}]
By Proposition~\ref{lem:nash1}, when $K=1$ {\sc simple} achieves $\frac{e-1}{e}$ fraction of {\sc optimum} social welfare. Therefore the welfare in the equilibrium for $K=1$ is at least $\frac{e}{2(e-1)}\frac{e-1}{e} = \frac{1}{2}$ fraction of {\sc optimum} welfare. As {\sc optimum} is by definition independent of $K$, it suffices to show that the welfare of the Nash equilibrium cannot decrease as $K$ increases. {We show this in two steps. Full proof is in Appendix \ref{sec:app-comp}. As a first step,}  we  compare the welfare of {\sc simple} 
 with the welfare achieved in the equilibrium of a game in which high doctors can submit a list of length $K>0$ and low ones submit a list of length $1$. To see why this {also satisfies the bound of Proposition \ref{lem:nash1}}, note that  as $K$ increases, we will show in Proposition~\ref{prop:X_H} that
high tier doctors have more applications on top. This will certainly increase the welfare of the high doctors. As shown in Lemma \ref{lm:poa2},  this increase will at least compensate for the loss incurred by low tier doctors.

{Next, we show in Lemmas \ref{lm:poa3} and \ref{lm:poa4} that social welfare of this compared to {\sc simple} is further improved when low doctors also have a list of size $K$. The two lemmas break the proof into two cases, if all doctors send at least one application to the low tier, then welfare is at least as high as the {\sc simple} benchmark. In the case, when doctors send less than one application to the low tier, we use the equilibrium properties to show that welfare at this equilibrium is at least as high as the welfare when only the high tier could send $K>1$ applications.}
\endproof

\paragraph{Remark.} It is important to note that the result in Theorem~\ref{thm:K>1} is indeed dependent on our assumption that the number of high tier hospitals is the same as high tier doctors. The quality of Nash equilibria can be a lot lower than the optimum when there is a shortage of high tier hospitals. For an example, assume $v$ is huge, we have $n$ top tier doctors, and only $\alpha n$ top hospitals $\alpha<<1$, $(1-\alpha)n$ low tier hospitals, and $r=0$.  Now all top tier doctors applicants apply only to high tier hospitals if say $v>1/\alpha$ achieving overall welfare less than $2 v \alpha n$. On the other hand, the optimal social welfare is $2v\alpha n+(1-\alpha)(v+1)\approx v(1+\alpha)n$, so the ratio  is $ 2\alpha/(1+\alpha) \to 0$ as $\alpha \to 0$.



\section{Properties of Equilibria}\label{sec:empirical}
{In this section we study how the equilibrium strategies and the welfare achieved at the equilibrium are affected by the primitives of the market. While a few {properties} we can prove formally, for many we rely on simulations.}
{The basic parameters used in our simulations are}
\begin{itemize}
\setlength{\itemsep}{0pt}\setlength{\parsep}{0pt}\setlength{\parskip}{0pt}
\item The length of the preference lists $K$ varies between 1 and 10.
\item The number of doctors and hospitals of the same tier agree. However, we vary the proportion of high to low agents, by considering the proportion of high doctors, $1/(r+1)$ to be equal to $0.1, 0.2, 0.3, 0.4, 0.5$.
	\item As usual, the value of the low type is assumed to be equal to 1. For the value of the high type, we consider $v= 1.001, 1.01, 1.05, 1.1, 1.25, 1.5$, and all the way up to $v=10$ using $.5$ increments.
\end{itemize}

The simulation results can be found in Appendix~\ref{app:simul}, but we describe our main findings next.

\paragraph{Effect on Equilibrium Strategies.}
First we consider the equilibrium strategy of high doctors as a function of $K$, the number of applications, and a function of $r$, the ratio of low and high quality doctors and hospitals. We defer the proof to Appendix~\ref{app-proofs-xh}.

\begin{proposition}\label{prop:X_H}
Let $X_H^*(K,v,r)$ denote the equilibrium strategy for the high type doctors when the length of lists is $K$, the ratio of low to high doctors is $r$, and the ratio of values of high and low match is $v$, as we defined in Section \ref{sec:eq}. Then, $X_H^*(K,v,r)$ is a monotone increasing function of $K$, $v$, and decreasing in $r$, e.g., the number of applications from the high doctors to high hospitals strictly increases with the length of the list.
\end{proposition}

\proof[Sketch]
Let the equilibrium strategy be $X^*=X^*(K)$ with the upper integer part $k=\lceil X^* \rceil$. Consider the solution where with $K+1$ applications, doctors still send $X^*$ applications to high hospitals. We show that doctors sending $k$ applications on top are better off than doctors sending $k-1$ applications. This shows that in equilibrium there must be more applications on top implying $X^*(K+1) \geq X^*(K)$.
Using similar arguments, one can show the dependence on both $v$ and $r$.
\endproof

{One would expect the equilibrium strategy to increase}
``smoothly" in $K$: if one extra slot is available ($K$ increases by one), the number of high hospitals ranked will increase by at most one. {Surprisingly,} 
this is not true in general. As an example,  consider an instance with
the same proportion of high and low doctors and high and low hospitals and $v=1.001$. Then, when $K=5$, the high-type doctors play the pure strategy $(1,4)$, and when $K=6$ they play $(3,3)$.
This discontinuity can be {understood} 
as follows: as $X^*_H$ is increasing in $K$, by adding an extra slot to the list more high-type doctors will match with high-type hospitals (on average). This implies that the availability of low type hospitals increased as less high-type doctors will attempt to match to them; thus, a doctor needs to list less low-type hospitals to obtain the same probability of being matched to one of them, potentially devoting more space to listing high-type hospitals. Although not true in general, we tested it computationally and it appears that if $v$ is ``big enough" then $X_H^*(K)<X_H^*(K+1)\leq X_H^*(K)+1$.
%

On the other hand, the effect that a change in primitives has in the strategy for low doctors $X^*_L$ is more involved.
For instance, one would like to establish a monotonicity result in $K$ for the strategy of the low-type. However, both situations might arise and the strategy of the low-type doctors can either increase or decrease. This is maybe not so surprising: as  $X^*_H$ is increasing in $K$, so more high-type doctors will match with high-type hospitals on expectation and thus less hospitals will be free. Hence, even though low doctors might use the extra space in their list to try to reach out to high hospitals, they are less likely to be accepted and thus the value of listing a high-hospital also decreases. In general, we found that when $v$ is relatively high, $X_L^*(K)$ is weakly increasing; even though as $K$ increases it is less likely that an application will be accepted, the fact that $v$ is high makes it attractive to spend some slots reaching to high hospitals. (This is not a surprise; if $v$ is sufficiently high, all high doctors apply high and thus they don't have a direct externality on the low doctors - low hospitals matches.) In addition, when $v$ is close to one, $X^*_L(K)$ is decreasing. This can be {understood} 
by the fact that, when $K$ is low, high doctors are very likely to be matched to low-type hospitals and thus low hospitals become less attractive to low doctors as they might already be taken. However, as $K$ increases, $X^*_H$ also increases, making high hospitals less attractive (more of them are taken) and low hospitals potentially more attractive as less high doctors might be matching low. Finally, for intermediate values of $v$ (e.g, $v=1.5$) we observe that $X^*_L$ might decrease and then increase or the other way around, depending on $r$. See Figure~\ref{fig:equil_low_type}.

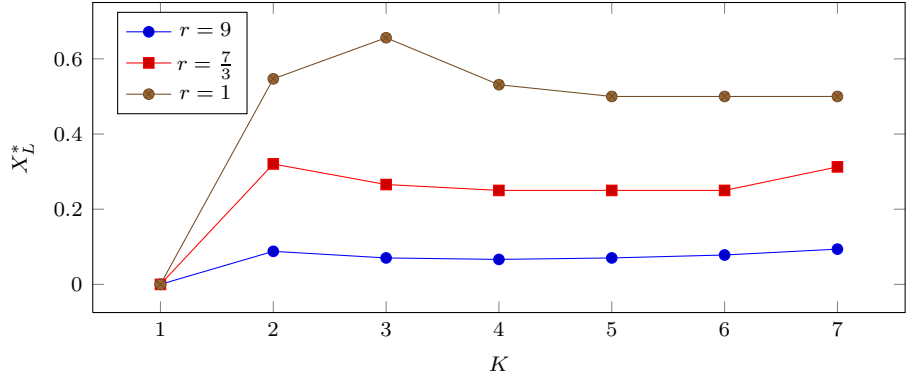
\begin{figure}\label{fig:equil_low_type}
\centering
\begin{tikzpicture}
\begin{axis}[
xlabel=$K$,
ylabel=$X^*_L$,
legend pos= north west,
x=1.5cm,y=5cm,
ymax=0.75
]
\addplot table {
1	0
2	0.087891
3	0.070312
4	0.066406
5	0.070312
6	0.078125
7	0.09375		
};
\addlegendentry{$r=9$}

\addplot table {
1	0
2	0.32031
3	0.26562
4	0.25
5	0.25
6	0.25
7	0.3125
};
\addlegendentry{$r=\frac{7}{3}$}

\addplot table {
1	0
2	0.54688
3	0.65625
4	0.53125
5	0.5
6	0.5
7	0.5
};
\addlegendentry{$r=1$}

\end{axis}
\end{tikzpicture}
\caption{Equilibrium strategy for the low type ($X^*_L$) as a function of the length of the list $K$ for different market compositions. In all cases, $ v=1.5$. }
\end{figure}

Regarding the dependence on $v$, when $K$ is low,  $X^*_L$ might either increase or decrease as $v$ increases. On the other hand, it always increases for high $K$. The intuition is similar that of the dependence on $K$; when $K$ is low, an increase in $v$ can have a huge impact on $X^*_H$, but when $K$ relatively high, most applications of high doctors will be to high hospitals anyway, so the impact in $X^*_H$ is less.  Finally, $X^*_L$ always increases as $r$ decreases. This is because of two effects, {both due to $X^*_H$ decreasing.} First, 
more high hospitals {remain} available to low doctors, which makes them more attractive. Second,
there are {relatively} more high doctors and less low hospitals, so the number of low hospitals available to low doctors decreases. 

\paragraph{Effect on Welfare.}
Regarding the social welfare produced by the Nash equilibrium solution we note that it is always increasing in $K$. In general, the number of within-tier matches increases as $K$ increases, which agrees with the intuition that if the doctors could submit the full-preference list, the assortative matching would arise. The welfare is also increasing in $v$.

Surprisingly, the welfare is not always increasing in the ratio of high to low doctors, even though the percentage of high agents for a fixed size population increased. In particular, when $v$ is close to one, the welfare decreases as a higher proportion of doctors and hospitals is of high type. This is due to the fact that high doctors split their applications between high and low hospitals; hence, it is more likely that low doctors will apply to a hospital that is already taken by a high doctor, thus wasting the applications and reducing the total welfare. 

\section{Conclusion}\label{sec:conclusion}
We considered a stylized model to study the incentive properties of the Gale-Shapley mechanism in systems with limited number of applications. As opposed to previous work, we do not assume that {the actual preference lists of agents is short, nor assume that} agents on both sides are (almost perfectly) vertically differentiated, but rather have a tier structure that allows us to have both horizontal and vertical differentiation. We characterized the equilibrium behavior in this game, {showed some surprising properties of the equilibria}, and used {the characterization} to derive bounds on the total welfare that can be achieved.

While this paper is a first step towards understanding how these markets operate, there are still many interesting open directions. First, it would be good to understand whether the insights change if more tiers are added. Second, although our tiered structure allows to accommodate for some horizontal and vertical differentiation, it would be good to understand whether the results could be extended when preferences are given by more {realistic} 
choice models. Finally, the bounds on the total welfare achieved by the different are not tight and could potentially be improved. We leave all these as future research.

\subsubsection*{Acknowledgement}
We want to express our deepest gratitude to Anna Karlin for her contributions to this project, and for many fun and insightful discussions.

\bibliographystyle{abbrv}
\bibliography{GSshort}

\begin{thebibliography}{10}
\setlength{\itemsep}{0pt}\setlength{\parsep}{0pt}\setlength{\parskip}{0pt}
\bibitem{Arnosti2015}
N.~Arnosti.
\newblock Centralized clearinghouse design: A quantity-quality tradeoff.
\newblock {\em Working Paper}, 2016.

\bibitem{Avery2010}
C.~Avery and J.~Levin.
\newblock Early admissions at selective colleges.
\newblock {\em The American Economic Review}, 100(5):2125--2156, 2010.

\bibitem{Chade2014}
H.~Chade, G.~Lewis, and L.~Smith.
\newblock Student portfolios and the college admissions problem.
\newblock {\em The Review of Economic Studies}, 81(3):971--1002, 2014.

\bibitem{Drummond2016}
J.~Drummond, A.~Borodin, and K.~Larson.
\newblock Natural interviewing equilibria for stable matching.

\bibitem{Gale1962}
D.~Gale and L.~S. Shapley.
\newblock College {A}dmissions and the {S}tability of {M}arriage.
\newblock {\em American Mathematical Monthly}, 69(1):9--15, 1962.

\bibitem{Gale1985}
D.~Gale and M.~Sotomayor.
\newblock Ms. machiavelli and the stable matching problem.
\newblock {\em The American Mathematical Monthly}, 92(4):261--268, 1985.

\bibitem{Hafalir2014}
I.~E. Hafalir, R.~Hakimov, D.~K{\"u}bler, and M.~Kurino.
\newblock College admissions with entrance exams: Centralized versus
  decentralized.
\newblock {\em Working Paper}, 2014.

\bibitem{ImmorlicaMahdian2005}
N.~Immorlica and M.~Mahdian.
\newblock Marriage, honesty, and stability.
\newblock In {\em Proceedings of the {S}ixteenth {A}nnual {ACM}-{SIAM}
  {S}ymposium on {D}iscrete {A}lgorithms}, 53--62. ACM, New
  York, 2005.

\bibitem{Kadam2015}
S.~V. Kadam.
\newblock Interviewing in matching markets.
\newblock {\em Working Paper}, 2015.

\bibitem{Kleinberg2016}
R.~Kleinberg, B.~Waggoner, and E.~G. Weyl.
\newblock Descending price optimally coordinates search.
\newblock In {\em Proceedings of the {ACM} Symposium of Ecomomic and
  Computation}. ACM, New York, 2016.

\bibitem{Roth1982}
A.~E. Roth.
\newblock The economics of matching: Stability and incentives.
\newblock {\em Mathematics of Operations Research}, 7(4):617--628, 1982.

\bibitem{Roth1992}
A.~E. Roth and M.~Sotomayor.
\newblock Two-sided matching.
\newblock {\em Handbook of game theory with economic applications}, 1:485--541,
  1992.


\end{thebibliography}

\bigskip
\pagebreak
\appendix
\section*{Appendix}\label{sec:appendix}
\section{Proof of Theorem~\ref{thm:equilibrium_two_tiers}}
\proof[of Theorem~\ref{thm:equilibrium_two_tiers}]
We divide the proof of Theorem~\ref{thm:equilibrium_two_tiers} into two lemmas as was explained in Section \ref{sec:eq}. (For ease of reading we repeat the statement of the lemmas here.)

\vspace{0.1cm}

\noindent
\textbf{Lemma \ref{lem:structure_two_tiers} [Structure of equilibria]}   In a market with two tiers, a symmetric equilibrium always exists and is unique, and for each type of agents $i\in \{L,H\}$, the equilibrium strategy is either pure, or is a randomization between consecutive strategies $(k_i, K-k_i)$ and $(k_i+1, K-k_i-1)$ for some $k_i\in \{0, \ldots, K\}$.

\proof We show that the structure is satisfied for the high doctors strategy. The proof for low doctors is essentially the same, but we must account for the probabilities $1-\alpha_H$ and $1-\alpha_L$, that a high and low hospital respectively are already taken by a high doctor (in that case, the low doctor gets automatically rejected).

Consider the market for high doctors and suppose that there is a symmetric equilibrium in which the probability that a doctor is accepted by a top hospital he applies to is $p$, and $p'$ be the probability that he is accepted by a bottom hospital that he applies to. These probabilities are as defined in Section~\ref{sec:model}.

Now consider the best response of a single player.  That player's goal is to choose $(y, K-y)$
to maximize his utility given what everyone else is doing. That is, he chooses (integer)
$y \in [0,K]$ to maximize
$$v(1- (1-p)^{y})+(1-p)^y(1- (1-p')^{K-y}) = v-(v-1)(1-p)^{y}-(1-p')^K \left(\frac{1-p}{1-p'}\right)^{y}.$$

Therefore, given the probabilities $p, p'$ we define  $f(y)$, the utility function of an agent when he lists $y$  hospitals of the top tier as follows:
\begin{align} \label{eq:response}
f(y) = v-(v-1)(1-p)^{y}-(1-p')^K \left(\frac{1-p}{1-p'}\right)^{y}
\end{align}

Note that, given $p$ and $p'$,  the best response of a single agent must maximize $f(y)$. (By the large market assumption, $p$ and $p'$ remain unchanged regardless of the agent's action.)
Note that the  function $f$ is concave when viewed as a function of a real variable $y$: It is of the form $c-a\alpha^y-b\beta^y$, where all of $a,b,c, \alpha, \beta>0$, $\alpha<1$, $\beta>1$.\footnote{Note that, at an equilibrium, we must have $p<p'$ as $v>1$.} The second derivative of this function is
$$
-a\ln^2(1/\alpha)\alpha^y-b\ln^2 \beta\beta^y
$$
which is negative for all $y$, so the function is strictly concave. Thus, the function either achieves its maximum over integers at
some integer $k$, or it is equal at two adjacent integer values (and strictly smaller at all other integer values).
\endproof

\noindent
\textbf{Lemma \ref{lem:existence_two_tiers} [Existence of equilibria]} A symmetric equilibrium always exists.

\proof
First, we show existence for the high ranked doctors. For a proposed solution $X:=k + x$ ($k=\lfloor X \rfloor$, as usual), with $x$ fraction of the top doctors listing $k+1$ high-type hospitals, and $K-k-1$ low-type hospitals, and the remaining $(1-x)$ fraction listing $(k, K-k)$ high-type/low-type hospitals respectively. The variable $X=k+x$ is the expected number of slots that a doctor uses to rank high-type hospitals. $X$ ranges in $[0,K]$.

Let $p(X)$ (resp. $p'(X)$) be the probability that {a single application of} a doctor is accepted by an individual top (resp. low-type) hospital he interviews with, when the strategy is given by $X$. In Claim \ref{cl:p_and_pprime} (stated after this proof) we show that $p(X)$ is continuous and monotone decreasing, and $p'(X)$ is continuous and monotone {increasing}. Intuitively, the more slots used to rank high hospitals, the smaller the probability that a single application succeeds in getting accepted, and this relation is continuous.  Clearly, $p'(K)=1$ and $p(0)=1$, as with no proposals sent there, the probability of a proposal being accepted is 1. If $vp(K)\ge 1$, then sending all proposals to the high-type hospitals is an equilibrium, as the added value of a proposal to the upper tier (conditioned on not yet being matched) is $v p(K)\ge 1$, more than the value of being accepted in a low hospital.

Now assume {$v p(K)< 1=p'(K)$}. Using some $\ell$ slots to rank low-type hospitals, a doctor has a $1-(1-p')^{\ell}$ probability of getting accepted. If an equilibrium agents exists where agents apply to either $k$ or $k+1$ high-type hospitals (by the above lemma, this are the only kind that can exists), then we must have that at the choice of using the $k+1$ slot for a high-type or a low-type hospital, the resident is neutral. Then, we need to have
$$
pv+(1-p)(1-(1-p')^{K-k-1})=(1-(1-p')^{K-k}).
$$

Consider the function $g(X)$ defined as the difference of the two sides:
\begin{equation}\label{eq:g_fn}
g(X):= pv+(1-p)(1-(1-p')^{K-k-1})-(1-(1-p')^{K-k}),
\end{equation}

\noindent with $k$ being the integer part of $X$, and $p, p'$ being $p(X), p'(X)$ respectively. When $X$ is $0$, then recall $p(0)=1$, and hence this value is positive. When $X=K$ then this value is negative, as $p(K)v< 1=p'(K)$. By continuity of $p, p'$, the function $g$ is continuous at fractional values of $X$ but can be discontinuous when $X$ is integer.

We first claim that, if $g(X)=0$ for any $X$ with integer part $k$, then this is an equilibrium. To see why note that the resident in this case is neutral between using $k$ or $k+1$ links to the upper part. Recall that the function $f(y)$ above expressing the value of using $y$ links up is strictly concave, so if two neighboring integers have equal value these are the maximum.


Next, suppose the two sides are never equal. We claim that then a pure strategy Nash equilibrium must exists. Recall that $g(0)>0$ and $g(K)<0$. Note $g$ is continuous from the right, but at integer points it is not continuous  from the left, as at that point the integer part of $X$ changes. If $g$ is never zero (both parts of the above equation are never equal), there must be an integer $X=k$ such that $g(k)<0$, but the lim-sup is positive when $X$ approaches $k$ from the bottom. We claim that, in this case, all doctors playing  $(k, K-k)$ is an equilibrium. As $g(k)<0$, sending $k+1$ applications to high-type hospitals is worse than sending $k$ such applications. Here we use that $g$ is decreasing in $X$, which follows from monotonicity of $p$ and $p'$. At the same time, if we consider a value $X<k$ but very close to $k$, $p$ and $p'$ are not affected (or barely), but $k$ got replaced by $k-1$, and thus the sign of $g$ changed, so now we get that listing $k$ high-type hospitals is better than listing $k-1$. Again the function expressing the value of best response is concave, so if $k$ is better than both $k-1$ and $k+1$, it must be the (unique) optimum.

{The same argument with some slight  modification holds for low tier doctors. Assume that the allocation of top tier doctors is fixed. For a proposed solution, consider same definition of $X, k, x, p $ and $p'$ for low tier doctors where $X$ is the expected number of applications sent to top tier hospitals. We need to consider $(1-\alpha_H)$ and $(1-\alpha_L)$, the fraction of high and low hospitals respectively that are spoken for by high doctors. Therefore $p'(K)$ and $p(0)$, respectively change to $\alpha_L$ and $\alpha_H$ in this case. Here if $vp(K)\ge \beta$, then sending all proposals to the high-type hospitals is an equilibrium, as the added value of a proposal to the upper type (conditioned on not yet being matched) is $v p(K)\ge \alpha_L$, more than the maximum value possible on the lower type. And for the other cases $v p(K)< \alpha_L=p'(K)$ same argument holds for low doctors. 
\endproof

\begin{claim}\label{cl:p_and_pprime}
Let $X \in [0,K]$ be a the expected number of top applications in the strategy or high type doctors. Then:
\begin{enumerate}
\item $p(X,r)$ is strictly decreasing in $X$ for all $r>0$.
\item $p'(X,r)$ is strictly increasing in $X$.
\end{enumerate}
\end{claim}

{The properties are easy to see for a fixed (finite) $n$, additional applications, can only decrease the probability that one is accepted. Since this is true for all finite $n$, its also true in the limit in our large market model.}

\label{sec:app_eq}
\section{Proofs on Efficiency of Equilibrium}

\paragraph{\textbf{Proof of Proposition~\ref{lem:nash1}.}}
As discussed in Section~\ref{sec:comp}, under the assumption of $v > e/(e-1)$, no high doctor  applies to a low tier hospital. Let $x$ be the fraction of low tier doctors that apply to the top tier in equilibrium. As discussed previously, social welfare is $2vn(1-1/e)+n(v+1)(1-e^{-xr})/e+2rn(1-e^{-(1-x)})$.
There are three different cases for $x$: $x=0$,  $0<x<1$ and $x=1$.
If $x=0$, the equilibrium welfare is same as {\sc simple}.
To show this for the case $0<x<1$, we are interested in finding a lower bound on the difference between the welfare at equilibrium and  {\sc simple} welfare. In equilibrium compared to {\sc simple} there are less matches to low tier hospitals, and more matches to top tier hospitals. Less matches to low tier hospitals causes a decrease in welfare equal to
$$DEC = 2rn\left[1-\frac{1}{e}-(1-e^{-(1-x)})\right].$$
The matches from low tier doctors to high tier hospitals cause an increase of
$$INC = (v+1)n\frac{1}{e}(1-e^{-xr}).$$
To bound decrease in welfare, we will lower-bound this increase by dropping the +1 from $(v+1)$.
Now substituting $v$ from the equilibrium condition above, we get that the increase is more than
$$n \dfrac{exr(1-e^{-(1-x)})}{(1-x)(1-e^{-xr})}\frac{1}{e}(1-e^{-xr})
=n\frac{xr(1-e^{-(1-x)})}{1-x}$$
The total difference ($DIFF$) is greater than:
$$nr\left( \frac{x(1-e^{-(1-x)})}{1-x} -2 \left( (1-\frac{1}{e}-(1-e^{-(1-x)}) \right) \right)
=nr\left( \frac{x(1-e^{-(1-x)})}{1-x} -2 \frac{e^x-1}{e} \right)$$

We claim that this function is monotone decreasing in $x$, and hence its infimum occurs when $x$ tends to $1$. To see that the function is decreasing, consider the derivative with respect to $x$, which is:
$$nr\frac{e-e^x(x^2-3x+3)}{e(1-x)^2}$$
and is indeed negative for all $0\leq x < 1$. The value of the difference with fixed $r$ and $n$ is lowerbounded by the limit as $x$ approaches 1, which is $rn(\frac{2-e}{e})$.

Recall that social welfare of {\sc simple} is $2n(r+v)(1-\frac{1}{e})$. The ratio of equilibrium  to {\sc simple} is at least:
$$\frac{SW(equilibrium)}{SW(\text{\sc simple})} \geq \frac{2n(r+v)(1-\frac{1}{e}) + rn\frac{2-e}{e}}{2n(r+v)(1-\frac{1}{e})} $$
$$= 1+\frac{rn\frac{2-e}{e}}{2n(r+v)(1-\frac{1}{e})} \geq 1+\frac{rn\frac{2-e}{e}}{2nr(1-\frac{1}{e})}=\frac{e}{2(e-1)} > 0.79 $$

Finally, note that when $K=1$, we have $x=1$ only if $v(1-e^{-r})/(er) \ge 1$. Then, if $x=1$, we have $DEC = 2rn(1-1/e)$ and $INC=(v+1)n(1-e^{-r})/e \ge rn$, so here the overall difference is also lower-bounded by $rn(\frac{2-e}{e})$. This gives the same bound as the case of $0<x<1$.

\endproof


\paragraph{\textbf{Proof of Theorem~\ref{thm:K>1}}} We will show in Lemmas~\ref{lm:poa3} and \ref{lm:poa4}  that the efficiency of an equilibrium is at least a $\frac{1}{2(e-1)}$ of the {\sc simple} benchmark. As shown previously {\sc simple} generates a $\frac{e-1}{e}$ fraction of the {\sc Optimum} welfare. Therefore social welfare of equilibrium is at least $\frac{1}{2(e-1)}\frac{e-1}{e}=\frac{1}{2}$ of {\sc Optimum}.

\begin{lemma}\label{lm:poa1} Suppose $v \ge \frac{e}{e-1}$ and the high tier is allowed to submit a list of length $K \geq 1$. Then, the social welfare resulting {from matches of} high tier doctors {with high tier hospitals} is not less than their social welfare when $K=1$.
\end{lemma}

\proof
With $v \ge \frac{e}{e-1}$ and $K=1$, everybody in high tier applies to high hospitals. Based on Proposition~\ref{prop:X_H} the number of applications to top is increasing in $K$. Therefore there will be more matched doctors to high tier. This implies that social welfare from high tier doctors is more than in the $K=1$ case.
\endproof

\begin{lemma}\label{lm:poa2}
Social welfare of equilibrium where high tier doctors have lists of size $K\ge 1$ and low tier  doctors have lists size $1$ is greater than or equal to $\frac{e}{2(e-1)}$ of the social welfare of {\sc simple}.
\end{lemma}

\proof
{By Proposition \ref{prop:X_H}}, the number of matches of high tier to high tier (HH) is higher with $K \geq 1$. Also there might be matches from high tier of doctors to low tier of hospitals (HL).  Suppose after high doctors' applications, $f$ fraction of high tier hospitals and $f'$ fractions of low tier hospitals are free.

Suppose that in equilibrium, $x$ fraction of low tier doctors apply to top tier and $1-x$ fraction apply to low tier. {At an equilibrium {with $0 <x<1$}, the probability of acceptance on top times $v$ must be equal to the probability of acceptance on low.}
$$v\frac{f(1-e^{-xr})}{xr}=\frac{f'(1-e^{-(1-x)})}{1-x}$$

The $DEC$ and $INC$ terms (as defined in proof of \ref{lem:nash1})  change to following quantities:
$$DEC = 2rn [1-\frac{1}{e}-f' \times (1-e^{-(1-x)})]$$
$$INC = (v+1)nf(1-e^{-xr}) + rn(v+1)(1-f') + 2(1-f-\frac{e-1}{e})$$

With fixed $f$ and $f'$ we would like to find a lower-bound on $\frac{INC-DEC}{SW(\text{\sc simple})}$.
Let $INC_{LH}$ be social welfare added by low to high matches. $INC-INC_{LH}$ is independent of $x$:
$$INC_{LH} \geq vnf(1-e^{-xr})$$
$$=\dfrac{f'(1-e^{-(1-x)})xr}{f(1-e^{-xr})(1-x)}nf(1-e^{-xr})$$
$$=\frac{nxrf'(1-e^{-(1-x)})}{1-x}$$

To find a lower-bound on $INC-DEC$, we first find a lower-bound on $INC_{LH}-DEC$.
$$INC_{LH}-DEC \geq nr \left( \dfrac{xf'(1-e^{-(1-x)})}{1-x} -2( 1-\frac{1}{e}-f'+f'e^{-(1-x)} )   \right)$$

By separating the part independent from $x$, we have:
$$=nr\left( \dfrac{xf'(1-e^{-(1-x)})}{1-x} -2f'e^{-(1-x)} \right) -2nr(1-\frac{1}{e}-f') $$

The derivative of this expression with respect to $x$ is negative, therefore its infimum happens when $x$ tends to 1.
$$INC_{LH} \geq -nrf' -2nr (1-\frac{1}{e}-f') = nr(f'+\frac{2}{e}-2)$$

The total difference ($DIFF$) is greater than:
$$rn(f'+\frac{2}{e}-2)+rn(v+1)(1-f)+2nv(-f-\frac{e-1}{e})$$
$$\geq rn(f'+\frac{2}{e}-2)+rn(1-f')$$

Therefore the ratio of welfare at equilibrium to welfare of {\sc simple} is at least:
\begin{eqnarray*}
\frac{SW(equilibrium)}{SW(\text{\sc simple})} &\geq & 1+ \frac{rn(f'+\frac{2}{e}-2)+rn(1-f')}{2n(r+v)(1-\frac{1}{e})} \\
& \geq & 1+\frac{rn\frac{2-e}{e}}{2nr(1-\frac{1}{e})} = \frac{e}{2(e-1)} > 0.79
\end{eqnarray*}
\endproof

\begin{lemma}\label{lm:poa3} Suppose $v > \frac{e}{e-1}$ and doctors have a list of $K \geq 1$, and suppose in equilibrium, doctors in low tier send at least one application to low tier hospitals. The social welfare of equilibrium in this case is not less than {\sc simple} benchmark.
\end{lemma}

\proof
Since the behavior of high doctors is independent of low doctors because of their priority, as discussed in Lemma~\ref{lm:poa1}, their social welfare from top tier hospitals increases and is at least their social welfare in {\sc simple} case.

Consider now the low tier doctors matches in equilibrium. Let us call this matching $M$. Note that we are not considering high tier doctors in this matching. We claim that social welfare by low tier doctors is at least their social welfare if we do not consider their applications to top tier. Let the matching by low tier doctors in this case be $M'$. The low tier doctors matched in $M'$ remain matched in $M$. Either they are matched to a high hospital, which contributes more to the social welfare, or if their applications on top all got rejected, they still have a spot in low tier hospital, because they had this spot with more competition when we ignored every low tier doctor matches to top. This satisfies the claim.

Now consider the high doctors matches in equilibrium and  the matching $M'$ for low doctors. We claim this has a higher social welfare than {\sc simple}. Consider high-to-high (HH), high-to-low (HL) and low-to-low (LL) matches. where the first letter refers to doctor's tier and the second to hospitals' one. HH links in equilibrium make more social welfare than HH links in {\sc simple}, due to {Lemma~\ref{lm:poa1}}.
Since every low tier doctor sends at least one application to the low tier in $M'$, the number of matches is at least $nr(1-1/e)(1-f)$ where $ f$ is the fraction of low tier hospitals occupied by high tier doctors. Therefore, the number of matches from $LL+HL$ is at least $nrf + nr(1-1/e)(1-f) \ge nr(1-1/e)$ and hence social welfare is more than {\sc simple}.
\endproof

\begin{lemma}\label{lm:poa4}
Suppose $v \ge \frac{e}{e-1}$ and doctors have a list of $K \geq 1$, and suppose in equilibrium, doctors in low tier  send 
{less than} one application to low tier hospitals {in expectation}. The social welfare of equilibrium in this case is better than equilibrium when top tier has list of size $K$ and low tier has list of size $1$.
\end{lemma}
\proof

Suppose that, for the low doctors, the expected number of applications  they send to the lower tier is less than 1. (In other words $X^* >K-1$, where $X^*$ is the expected number of application a low tier doctors sends to the high tier.) 
Based on 
equilibrium characterization, $pv { \ge} p'$, where $p$ is the success of a link from lower tier on top and $p'$ is the probability success of a link from low tier to top tier.

Consider the equilibrium when top hospitals have lists of size $K$ and low hospitals have lists of size $1$. Let $q$ and $q'$ be the success probability of an application from low tier doctors to high and low tier hospitals respectively. Call the equilibrium solution of this case $E'$, and the equilibrium solution with $K$ applications $E$.

{In equilibrium $E$ with low tier doctors having $K>1$ applications, and sending in expectation less than 1 application to low tier hospitals, they are sending more application to high tier, so the high tier success probability $p$ is less than $q$}. 

We claim that the number of LL applications in $E'$ is less than in $E$. Suppose the opposite, this means for utility of an application to low we have $q'<p'$ due to higher chance of acceptance with less number of overall applications. {Combining this with $p'\le vp$, and $p \le q$, we get $q' <p' \le vp \le vq$,
which implies $vq>q'$. In this case sending applications down is not equilibrium in $E'$,  which contradicts with the  assumption that LL applications in 
{$E$} is less than in 
{$E'$}. Therefore  there must be more applications from low tier to low tier in equilibrium for K rather than 1. Which means social welfare is higher for $K>1$.
}
\endproof

\label{sec:app-comp}
\section{Appendix to Section~\ref{sec:empirical}}
\subsection{Proof of Proposition~\ref{prop:X_H}}
\label{app-proofs-xh}

\proof We first prove that $X^*_H$ increases in $K$.
Suppose that $X^*(K)$ is not an integer.  Let $g(X,K)$ be defined as
$$
g(X,K):= pv+(1-p)(1-(1-p')^{K-k-1})-(1-(1-p')^{K-k}),
$$
where $k$ is the integer part of $X$, and $p$ and $p'$ is the acceptance probability of a single application with $X$ of the applications going to the top part in expectation. This definition of $g$ agrees with the one in Eq.~\eqref{eq:g_fn}, where we made the dependence on $K$ explicit. Here $$p''=(1-(1-p')^{K-k-1})$$ is the probability that $K-k-1$ applications sent to the lower tier results in success. Using this additional notation, we can write
$$
g(X,K):= pv+(1-p)p''-(p''+(1-p'')p')=p(v-p'')-p'(1-p'')
$$

As $X^*=X^*(K)$ is an equilibrium, we have $g(X^*, K)=0$. Recall that $g(X,K)$ is decreasing in $X$ for every  $K$. Hence, to show that $X^*(K+1)>X^*(K)$ it suffices to show that $g(X^*, K+1) \ge 0$. Let $\hat p'$ and $\hat p''$ denote the probabilities corresponding to $p'$ and $p''$ in the last expression for $g(X^*, K+1)$, and note that $p$ is unchanged, as its only effected by $X$ and not by $K$.

We claim that $\hat p'\le p'$, as with more applications to the lower part, each has a smaller change of succeeding.

Second, we also claim that $\hat p''\ge p''$. This is true, as with more applications to the bottom part, more hospitals in the bottom will secure an application, and hence overall more applicants will be accepted to some hospital.

This now implies that both $1-\hat p''\le 1-p''$ and $v-\hat p''\le v-p''$, but since $v>1$ we also know that
$$\frac{v-\hat p''}{1-\hat p''} \ge \frac{v-p''}{1-p''}$$
Combining these gives us the bound as
\begin{eqnarray*}
g(X^*,K+1)&:= & p(v-\hat p'')-\hat p'(1-\hat p'')\ge p(v-\hat p'')- p'(1-\hat p'')\\
&=& (1-\hat p'')(\frac{v-\hat p''}{1-\hat p''}p-p')\ge (1-\hat p'')(\frac{v-p''}{1-p''}p-p')\\
&=&\frac{1-\hat p''}{1- p''}((v-p'')p-(1-p'')p')=\frac{1-\hat p''}{1- p''}g(X^*,K)=0
\end{eqnarray*}

Now, to show that it is increasing in $v$, it suffices to note that $g(X)$ is strictly increasing in $v$. As $g$ is decreasing in $X$, this implies that if $X^*$ is an equilibrium when $v_H=v$ ---thus $g(X^*)=0$---, then for $v'>v$ we have $g(X^*)>0$, which implies that the new equilibrium for $v'$ must be greater than $X^*$.

{In addition, it is decreasing in $r$.  To see why, note that for a fixed strategy $X$, the probability $p$ of matching to a high hospital (as defined in Section~\ref{sec:model}) remains the same. However, $p'$ increases; although the proportion of high-doctors matched to high-hospitals is the same regardless of $r$, 
the number of low hospitals also increased when $r$ increased. This causes $p'$ to increase. Rewriting $g$ as $p(v-1) - (1-p')^{K-k-1}(p'-p)$, it is easy to see that this is a decreasing function of $p'$. This implies that, everything else fixed, as $p'$ increases, the $X^*$ such a that $g(X^*)=0$ will decrease and thus the equilibrium strategy decreases.}
\endproof

\subsection{Simulation Results}
\label{app:simul}

\begin{table}
\label{tb:outcomes}
	\centering {\scriptsize 	\begin{tabular}{ c c||c|c|c|c||c|c|c|c||c|c|c|c|c|c|}
			\cline{3-14}
			& & \multicolumn{4}{ c| }{$rH = 0.1,~ rL = 0.9$} & \multicolumn{4}{ c| }{$rH = 0.3,~ rL = 0.7$} & \multicolumn{4}{ c| }{$rH = 0.5,~ rL = 0.5$}\\ \hline
			\multicolumn{1}{ |c| }{$v_H$} & $K$ & $X^*_H$ & $X^*_L$  & $w(S)$ & $w(N)$ & $X^*_H$ & $X^*_L$  & $w(S)$ & $w(N)$ & $X^*_H$ & $X^*_L$  & $w(S)$ & $w(N)$\\ \hline	 
			\multicolumn{1}{ |c| }{1.001}  & 1 & 0.094 & 0.095 & 1.264 & 1.223 & 0.252 & 0.266 & 1.265 & 1.163 & 0.402 & 0.430 & 1.265 & 1.132 \\
			\multicolumn{1}{ |c| }{1.001} & 2 & 0.125 & 0.143 & 1.544 & 1.521 & 0.344 & 0.445 & 1.544 & 1.500 & 0.609 & 0.781 & 1.544 & 1.505 \\
			\multicolumn{1}{ |c| }{1.001} & 3 & 1.000 & 0.000 & 1.681 & 1.667 & 0.500 & 0.438 & 1.682 & 1.644 & 0.688 & 1.031 & 1.682 & 1.650 \\
			\multicolumn{1}{ |c| }{1.001} & 4 & 2.000 & 0.000 & 1.763 & 1.759 & 1.000 & 0.109 & 1.764 & 1.712 & 0.750 & 1.188 & 1.764 & 1.724 \\
			\multicolumn{1}{ |c| }{1.001} & 5 & 3.000 & 0.000 & 1.817 & 1.816 & 2.000 & 0.000 & 1.817 & 1.781 & 1.000 & 0.813 & 1.818 & 1.768 \\
			\multicolumn{1}{ |c| }{1.001} & 6 & 4.000 & 0.000 & 1.855 & 1.855 & 4.000 & 0.000 & 1.855 & 1.853 & 3.000 & 0.000 & 1.855 & 1.812 \\
			\multicolumn{1}{ |c| }{1.001} & 7 & 5.000 & 0.000 & 1.882 & 1.883 & 5.000 & 0.000 & 1.883 & 1.883 & 4.000 & 0.000 & 1.883 & 1.857 \\ \hline
			\multicolumn{1}{ |c| }{1.01}  & 1 & 0.109 & 0.092 & 1.266 & 1.224 & 0.264 & 0.260 & 1.268 & 1.167 & 0.410 & 0.426 & 1.271 & 1.138  \\
			\multicolumn{1}{ |c| }{1.01} & 2 & 0.500 & 0.945 & 1.545 & 1.529 & 0.500 & 0.313 & 1.548 & 1.509 & 0.625 & 0.750 & 1.551 & 1.512  \\
			\multicolumn{1}{ |c| }{1.01} & 3 & 1.000 & 0.000 & 1.683 & 1.668 & 1.000 & 0.031 & 1.686 & 1.641 & 0.750 & 0.875 & 1.690 & 1.657  \\
			\multicolumn{1}{ |c| }{1.01} & 4 & 2.000 & 0.000 & 1.765 & 1.760 & 2.000 & 0.000 & 1.768 & 1.749 & 2.000 & 0.000 & 1.772 & 1.726  \\
			\multicolumn{1}{ |c| }{1.01} & 5 & 4.000 & 0.000 & 1.819 & 1.823 & 3.000 & 0.000 & 1.822 & 1.815 & 3.000 & 0.000 & 1.826 & 1.805 \\
			\multicolumn{1}{ |c| }{1.01} & 6 & 5.000 & 0.000 & 1.856 & 1.860 & 4.000 & 0.000 & 1.860 & 1.858 & 4.000 & 0.000 & 1.864 & 1.854  \\
			\multicolumn{1}{ |c| }{1.01} & 7 & 6.000 & 0.000 & 1.884 & 1.887 & 5.000 & 0.000 & 1.888 & 1.888 & 5.000 & 0.000 & 1.891 & 1.887  \\ \hline
			\multicolumn{1}{ |c| }{1.1}  & 1 & 0.258 & 0.073 & 1.277 & 1.239 & 0.381 & 0.203 & 1.302 & 1.208 & 0.504 & 0.324 & 1.328 & 1.202  \\
			\multicolumn{1}{ |c| }{1.1}  & 2 & 1.000 & 0.000 & 1.559 & 1.559 & 1.000 & 0.004 & 1.590 & 1.579 & 1.000 & 0.242 & 1.621 & 1.615  \\
			\multicolumn{1}{ |c| }{1.1}  & 3 & 2.000 & 0.000 & 1.698 & 1.702 & 2.000 & 0.000 & 1.732 & 1.739 & 2.000 & 0.000 & 1.765 & 1.763  \\
			\multicolumn{1}{ |c| }{1.1} & 4 & 3.000 & 0.000 & 1.781 & 1.786 & 3.000 & 0.000 & 1.816 & 1.827 & 3.000 & 0.000 & 1.851 & 1.860  \\
			\multicolumn{1}{ |c| }{1.1}  & 5 & 4.000 & 0.000 & 1.835 & 1.840 & 4.000 & 0.000 & 1.871 & 1.883 & 4.000 & 0.000 & 1.908 & 1.920  \\
			\multicolumn{1}{ |c| }{1.1}  & 6 & 5.000 & 0.000 & 1.873 & 1.877 & 5.000 & 0.000 & 1.910 & 1.921 & 5.000 & 0.000 & 1.947 & 1.960  \\
			\multicolumn{1}{ |c| }{1.1}  & 7 & 6.000 & 0.000 & 1.901 & 1.904 & 6.000 & 0.000 & 1.938 & 1.948 & 6.000 & 0.000 & 1.976 & 1.988  \\ \hline
			\multicolumn{1}{ |c| }{1.5}  & 1 & 0.883 & 0.000 & 1.328 & 1.320 & 0.895 & 0.000 & 1.454 & 1.433 & 0.914 & 0.000 & 1.580 & 1.552  \\
			\multicolumn{1}{ |c| }{1.5} & 2 & 1.000 & 0.088 & 1.621 & 1.636 & 1.094 & 0.320 & 1.775 & 1.810 & 1.328 & 0.547 & 1.930 & 1.956  \\
			\multicolumn{1}{ |c| }{1.5} & 3 & 2.000 & 0.070 & 1.765 & 1.782 & 2.000 & 0.266 & 1.933 & 1.980 & 2.000 & 0.656 & 2.102 & 2.173  \\
			\multicolumn{1}{ |c| }{1.5} & 4 & 3.000 & 0.066 & 1.851 & 1.866 & 3.000 & 0.250 & 2.028 & 2.069 & 3.000 & 0.531 & 2.204 & 2.266  \\
			\multicolumn{1}{ |c| }{1.5} & 5 & 4.000 & 0.070 & 1.908 & 1.920 & 4.000 & 0.250 & 2.089 & 2.125 & 4.000 & 0.500 & 2.271 & 2.325  \\
			\multicolumn{1}{ |c| }{1.5} & 6 & 5.000 & 0.078 & 1.947 & 1.958 & 5.000 & 0.250 & 2.133 & 2.163 & 5.000 & 0.500 & 2.318 & 2.364  \\
			\multicolumn{1}{ |c| }{1.5} & 7 & 6.000 & 0.094 & 1.976 & 1.986 & 6.000 & 0.313 & 2.164 & 2.192 & 6.000 & 0.500 & 2.352 & 2.392  \\ \hline
			\multicolumn{1}{ |c| }{3} & 1 & 1.000 & 0.121 & 1.517 & 1.512 & 1.000 & 0.350 & 2.023 & 2.010 & 1.000 & 0.572 & 2.529 & 2.519  \\
			\multicolumn{1}{ |c| }{3} & 2 & 2.000 & 0.215 & 1.852 & 1.860 & 2.000 & 0.875 & 2.470 & 2.489 & 2.000 & 1.000 & 3.087 & 3.179  \\
			\multicolumn{1}{ |c| }{3} & 3 & 3.000 & 0.289 & 2.017 & 2.026 & 3.000 & 1.000 & 2.690 & 2.728 & 3.000 & 1.000 & 3.362 & 3.449  \\
			\multicolumn{1}{ |c| }{3} & 4 & 4.000 & 0.352 & 2.116 & 2.124 & 4.000 & 1.000 & 2.821 & 2.862 & 4.000 & 2.000 & 3.526 & 3.573  \\
			\multicolumn{1}{ |c| }{3} & 5 & 5.000 & 0.406 & 2.180 & 2.187 & 5.000 & 1.000 & 2.907 & 2.945 & 5.000 & 2.000 & 3.634 & 3.685  \\
			\multicolumn{1}{ |c| }{3} & 6 & 6.000 & 0.469 & 2.225 & 2.231 & 6.000 & 1.250 & 2.967 & 2.993 & 6.000 & 2.000 & 3.709 & 3.757  \\
			\multicolumn{1}{ |c| }{3} & 7 & 6.875 & 0.531 & 2.258 & 2.264 & 6.875 & 1.500 & 3.011 & 3.035 & 6.875 & 2.750 & 3.764 & 3.802  \\ \hline
			\multicolumn{1}{ |c| }{10} & 1 & 1.000 & 0.508 & 2.402 & 2.342 & 1.000 & 1.000 & 4.678 & 4.889 & 1.000 & 1.000 & 6.953 & 7.600 \\
			\multicolumn{1}{ |c| }{10} & 2 & 2.000 & 1.000 & 2.933 & 2.915 & 2.000 & 1.281 & 5.712 & 6.015 & 2.000 & 2.000 & 8.490 & 8.785 \\
			\multicolumn{1}{ |c| }{10} & 3 & 3.000 & 1.000 & 3.194 & 3.232 & 3.000 & 2.000 & 6.220 & 6.413 & 3.000 & 2.078 & 9.246 & 9.712 \\
			\multicolumn{1}{ |c| }{10} & 4 & 4.000 & 1.000 & 3.350 & 3.396 & 4.000 & 2.000 & 6.523 & 6.726 & 4.000 & 3.000 & 9.697 & 10.019  \\
			\multicolumn{1}{ |c| }{10} & 5 & 5.000 & 1.156 & 3.452 & 3.484 & 5.000 & 3.000 & 6.722 & 6.809 & 5.000 & 3.688 & 9.993 & 10.212  \\
			\multicolumn{1}{ |c| }{10}  & 6 & 6.000 & 1.438 & 3.524 & 3.541 & 6.000 & 3.000 & 6.862 & 6.960 & 6.000 & 4.000 & 10.199 & 10.404  \\
			\multicolumn{1}{ |c| }{10}  & 7 & 7.000 & 1.688 & 3.576 & 3.587 & 7.000 & 3.000 & 6.963 & 7.058 & 7.000 & 4.750 & 10.351 & 10.493  \\ \hline
		\end{tabular}}\caption{Equilibria and welfare outcomes for different combinations of values $v_H$ and length of list $K$. We assume that the number of high hospitals and doctors agrees, as well as the number low doctors and low hospitals. However, we assume three different market compositions: the proportion of high doctors ($r_H$) is 0.1, 0.3, and 0.5; the proportion of low doctors ($r_L$) is always $1-r_H$).}
	\end{table}
	
\begin{table}
\centering
\label{tb:simpleVSnash}
\begin{tabular}{|c|c|c|c|c|}
\hline
$X^*_H$	&	$X^*_L$   & r  & v           & w(S)/w(N) \\ \hline
0.094	&	0.095		& 1		& 1.001 				&	1.117491166\\
1	&	0.3	&	10	&	6.171953564	&	1.003290867	\\	
1	&	0.4	&	10	&	8.328930557	&	1.01088794	\\	
1	&	0.5	&	10	&	10.76816087	&	1.018342694	\\	
1	&	0.6	&	10	&	13.47584871	&	1.025452035	\\	
1	&	0.7	&	10	&	16.45401694	&	1.032211033	\\	
1	&	0.8	&	10	&	19.71625006	&	1.038664643	\\	
1	&	0.9	&	10	&	23.28395803	&	1.04486198	\\	
1	&	0.1	&	1000	&	179.2345456	&	1.007484894	\\	
1	&	0.2	&	1000	&	374.2197676	&	1.014518933	\\	
1	&	0.3	&	1000	&	586.4670089	&	1.02102172	\\	
1	&	0.4	&	1000	&	817.6380872	&	1.027133691	\\	
1	&	0.5	&	1000	&	1069.560558	&	1.032951674	\\	
1	&	0.6	&	1000	&	1344.244542	&	1.038544177	\\	
1	&	0.7	&	1000	&	1643.901282	&	1.043960656	\\	
1	&	0.8	&	1000	&	1970.9636	&	1.049237358	\\	
1	&	0.9	&	1000	&	2328.108456	&	1.054401137	\\	\hline

\end{tabular}
\caption{Ratio between social welfare of Simple and Equilibrium with different parameters for $v$ and $r$ and $K=1$, showing that social welfare of Simple are in cases higher than equilibrium. The first row shows this phenomenon with $v$ close to $1$, while other instances have rather high $v$.}
\end{table}


\end{document}